\documentclass[
 reprint,
 superscriptaddress,
 preprintnumbers,
 amsmath
 amssymb,
 prx,
 floatfix,
]{revtex4-2}

\DeclareMathAlphabet{\mathsfit}{T1}{\sfdefault}{\mddefault}{\sldefault}
\SetMathAlphabet{\mathsfit}{bold}{T1}{\sfdefault}{\bfdefault}{\sldefault}

\usepackage{amsmath}
\usepackage{graphicx}
\graphicspath{{fig/}}   
\usepackage{dcolumn}
\usepackage{bm}
\usepackage{makecell}
\usepackage{braket} 
\usepackage{siunitx}
\usepackage{color}
\usepackage{array}
\usepackage{braket}
\usepackage{siunitx}
\usepackage{xr}
\usepackage{upgreek}
\usepackage[colorlinks]{hyperref}
\usepackage[capitalize]{cleveref}
\usepackage[caption=false]{subfig} 
\usepackage{multirow} 
\usepackage[T1]{fontenc}

\begin{document} 

\title{Device characterization of Si/SiGe double quantum dots \\ using exchange oscillations in Earth's magnetic field}

\def\RLEaffil{Research Laboratory of Electronics, Massachusetts Institute of Technology, Cambridge, MA 02139, USA}
\def\LLaffil{Lincoln Laboratory, Massachusetts Institute of Technology, Lexington, MA 02421, USA}
\def\Physaffil{Department of Physics, Massachusetts Institute of Technology, Cambridge, MA 02139, USA}
\def\EECSaffil{Department of Electrical Engineering and Computer Science, Massachusetts Institute of 
Technology, Cambridge, MA 02139, USA}
\def\UNSWaffil{School of Electrical Engineering and Telecommunications, The University of New South Wales, Sydney, New South Wales 2052,
Australia}
\def\Diraqaffil{Diraq, Sydney, New South Wales, Australia}
\def\ETHaffil{Department of Physics, ETH Zurich}

\author{Holly~G.~Stemp}
\altaffiliation{These authors contributed equally to this work.}
\affiliation{\RLEaffil}

\author{Harry~Hanlim~Kang}
\altaffiliation{These authors contributed equally to this work.}
\affiliation{\RLEaffil}
\affiliation{\EECSaffil}

\author{Chih~Hwan~Yang}
\affiliation{\UNSWaffil}
\affiliation{\Diraqaffil}

\author{Gabriel~D.~Cutter}
\affiliation{\RLEaffil}
\affiliation{\EECSaffil}

\author{Frederike~Brockmeyer}
\altaffiliation{Present address: Department of Physics, ETH Zürich, 8093 Zürich Switzerland}
\affiliation{\RLEaffil}
\affiliation{\EECSaffil}

\author{Patrick~J.~Strohbeen}
\affiliation{\RLEaffil}

\author{Max~Hays}
\affiliation{\RLEaffil}

\author{Jeffrey~A.~Grover}
\affiliation{\RLEaffil}

\author{William~D.~Oliver}
\email{william.oliver@mit.edu}
\affiliation{\RLEaffil}
\affiliation{\EECSaffil}
\affiliation{\Physaffil}

\date{\today}

\begin{abstract} 

Exchange-based semiconductor qubits encompass a broad family of encodings constructed from singlet- and triplet-like spin states, several of which are compatible with operation at zero applied magnetic field. Their reliable operation requires characterization of environmental noise, residual idle interactions, and exchange-dependent decay, but this characterization often relies on multi-axis control calibration or deliberately engineered magnetic-field gradients. A simpler zero-applied-field diagnostic is particularly valuable for hybrid semiconductor-superconductor systems, in which magnetic fields can degrade superconducting components. Here, we use the intrinsic magnetic-field gradient produced by residual nuclear spins in isotopically enriched Si/SiGe to implement exchange oscillations between two quantum dots as a characterization tool without a micromagnet, dynamic nuclear polarization, or prior multi-axis calibration. Using Carr-Purcell-Meiboom-Gill exchange sequences, we extend the singlet coherence from $T_2^*=1.17\pm0.02~\unit{\micro\second}$ to $T_2^{\mathrm{CPMG}}=74.8\pm1.8~\unit{\micro\second}$ with $N=70$ refocusing pulses. The oscillation phase resolves residual exchange in the tens-of-kilohertz regime and enables it to be mapped across the $(1,1)$ charge cell. These results establish intrinsic-gradient exchange oscillations as a simple, more relevant zero-field diagnostic for exchange-only and related semiconductor qubit encodings that is amenable to rapid, high-throughput device characterization.

\end{abstract}

\maketitle

\section{\label{sec:1} Introduction}
Electron spins confined in Si/SiGe quantum dots are promising building blocks for quantum processors \cite{burkard2023semiconductor, xue2022quantum, philips2022universal, hrl2026digitally}. Their small physical footprint and compatibility with established semiconductor fabrication techniques offer a compelling path towards the dense arrays required to integrate large numbers of qubits on a single chip \cite{ha2021flexible, george202412, neyens2024probing}.

Exchange-based semiconductor qubits encompass a broad family of qubit encodings \cite{divincenzo2000universal, medford2013self, petta2005coherent, shim2016charge, foulk2025singlet}. Two prevalent examples are singlet-triplet qubits, encoded in two spins in a double quantum dot, and exchange-only qubits, encoded in three spins in a triple quantum dot. In both cases, information is stored in collective spin states with singlet-like and triplet-like character, providing resilience to common-mode magnetic noise \cite{burkard2023semiconductor}. Their control strategies differ: singlet-triplet qubits combine a tunable exchange interaction with a difference between the local magnetic fields experienced by the two electrons to achieve two orthogonal control axes, whereas exchange-only qubits generate logical rotations by controlling multiple pairwise exchange interactions.

Despite these different control strategies, both architectures depend sensitively on the behavior of their constituent exchange-coupled spin pairs. Reliable operation therefore requires characterization of the coherence, residual exchange during idle periods, and decay processes associated with each exchange interaction. In exchange-only devices, extracting these properties through the encoded-qubit response requires prior calibration of multiple exchange-control axes. A simpler diagnostic that isolates an individual exchange-coupled pair can therefore complement the calibration of the complete logical-control space.

For conventional singlet-triplet control, the required magnetic-field gradient is commonly produced using an integrated micromagnet \cite{wu2014two, song2024coherence, jang2020individual} or dynamic nuclear polarization \cite{dial2013charge, foletti2009universal, bluhm2010enhancing, shulman2014suppressing, cerfontaine2020closed, barthel2012relaxation}. Dynamic nuclear polarization requires a sufficiently large and controllable nuclear-spin ensemble and is therefore less readily implemented in isotopically enriched materials. Micromagnets provide a more reproducible gradient but add fabrication requirements and are typically activated in the presence of an applied magnetic field. These requirements are particularly restrictive for hybrid semiconductor-superconductor architectures \cite{dijkema2025cavity, landig2019virtual, frey2012dipole, mi2018coherent, harvey2022coherent, kang2025remote}, in which magnetic fields can increase loss and degrade the performance of superconducting resonators and qubits \cite{kroll2019magnetic}. Methods that can characterize semiconductor spin devices without an intentionally applied magnetic field are therefore desirable for hybrid integration.

Even at zero applied magnetic field, however, a small intrinsic magnetic-field gradient remains in isotopically enriched Si/SiGe due to the hyperfine coupling between residual spinful nuclei and the confined electrons \cite{struck2020low, eng2015isotopically, witzel2012nuclear}. In this work, we exploit this intrinsic gradient to drive exchange oscillations and characterize a Si/SiGe double quantum dot at zero applied magnetic field \cite{petta2005coherent, maune2012coherent, berritta2024real, laird2006effect}. Although the gradient is small and fluctuates randomly between experimental repetitions, exchange refocusing pulses convert its aggregated effect into a robust singlet-return response.

We use these exchange oscillations to probe otherwise difficult-to-resolve properties of a Si/SiGe double quantum dot: qubit coherence during free-evolution, exchange coupling during idling and the decay rate of the singlet-state as a function of exchange interaction strength. Carr-Purcell-Meiboom-Gill (CPMG) sequences containing up to $N=70$ exchange refocusing pulses extend the singlet coherence time from $T_2^*=1.17\pm0.02~\unit{\micro\second}$ to $T_2^{\mathrm{CPMG}}=74.8\pm1.8~\unit{\micro\second}$. Within the same multipulse framework, the phase response of sequences containing different numbers of exchange pulses resolves residual exchange coupling in the tens-of-kilohertz regime and enables its value to be compared at representative locations within the $(1,1)$ charge cell. Measurements of the singlet-return decay under a constant exchange interaction additionally reveal a broad enhancement of the decay rate between exchange frequencies of approximately 20 and 40~MHz, providing a probe of exchange-dependent relaxation or leakage.

Together, these measurements establish intrinsic-gradient exchange oscillations as a simple characterization primitive for exchange-based semiconductor qubits, more relevant to the low-field environments in which exchange only qubits are typically operated. Because the protocol uses dephasing itself to generate the measurable singlet-return response, rather than relying on a specific microscopic origin of that dephasing, other zero-field dephasing mechanisms that are refocusable by exchange pulses could be used in the same way, making the method broadly applicable. Its operation without an intentionally applied magnetic field also makes it particularly well suited to quantum dots intended for integration with field-sensitive superconducting resonators and qubits.

\section{\label{sec:1} Exchange oscillations with nuclear field gradients}
    \begin{figure*}[ht]
            \centering\includegraphics[width=1\textwidth]{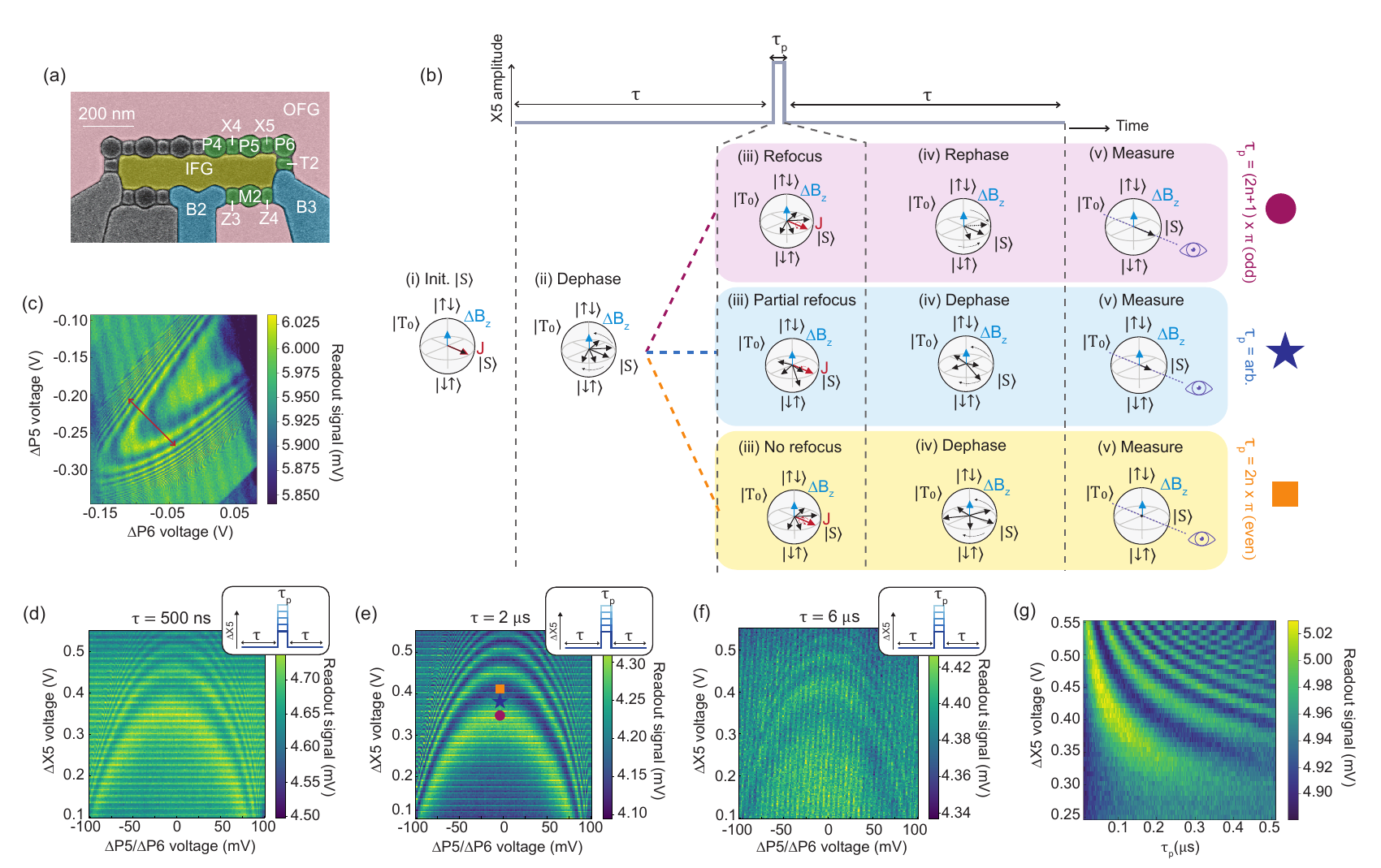} 
            \caption{\textbf{Exchange oscillations at zero magnetic field.} (a) False colored SEM image of a 1 x 6 SLEDGE Si/SiGe quantum dot device nominally identical to the device used in this work. Only the P5 and P6 dots in the right-hand triple quantum dot were operated. Pink gates represent the outer field gate, OFG, which is DC biased in depletion mode for quantum dot confinement. The yellow gate is the inner field gate, IFG, also DC biased in depletion mode for lateral dot confinement. Blue gates represent bath gates, which are DC biased in accumulation mode, to provide a supply of electrons to the quantum dots from the ohmic reservoirs (not shown). Green gates represent gates that are connected to both a DC and fast line, to allow for fast pulsing. Z3, M2 and Z4 gates make up the dot charge sensor. In this work, electrons are loaded into the quantum dots below gates P5 and P6, with X5 controlling the barrier between the dots and T2 controlling the barrier between the dots and the bath.  (b) Three distinct cases of the spin state evolution for the pulse sequence used to obtain the exchange oscillation fingerprints in (d-f). The sequence involves first initializing a singlet state in the (0,2) charge cell. We then wait at the idling point in the (1,1) charge cell for a time $\tau$, where the spin state undergoes dephasing. This is followed by a pulse on the X5, P5 and P6 gates for a time $\tau_{\text{p}}$, to increase the exchange interaction to a high value. We then wait for a final free-evolution time, $\tau$ in the (1,1) cell. For the case of $\tau_{\text{p}} = (2n+1) \times \pi$, where n is an integer, (purple circle) the exchange pulse provides a refocusing effect, resulting in the state returning to the singlet state at the end of the sequence. For the case of an arbitrary $\tau_{\text{p}}$ value (blue star), the exchange pulse produces a partial refocusing effect. For the case of $\tau_{\text{p}} = 2n \times \pi$, the exchange interaction provides no refocusing effect and thus the final state at the end of the sequence is a completely mixed state. The positions at which these three cases occur in the exchange fingerprint plot is shown by the corresponding symbols in (e). (c) $J_{\text{X5}}$ nonequilibrium cell showing the exchange oscillations between the electrons in quantum dots P5 and P6 as a function of P6 and P5 voltages, with $\Delta$X5 = 0.55 V. Red arrow denotes the detuning axis, along which the scans in (d-f) were taken. (d,e,f) Exchange fingerprint scans showing the exchange oscillations as a function of both the barrier, $\Delta$X5, voltage and the $\Delta$P6, $\Delta$P5 voltages with free-evolution times of (d), $\tau = 500$ ns, (e), $\tau = 2$~\unit{\micro\second} and (f), $\tau = 6$~\unit{\micro\second}. The center of the scan is chosen from the center of the detuning axis shown with the red arrow in (c). Pulse sequences in the top right of each plot denote that we are sweeping the amplitude of the exchange pulse in these scans. (g) Exchange oscillations between the electrons in P5 and P6 quantum dots as a function of $\Delta$X5 amplitude and pulse time. Note that P5 and P6 were kept fixed for the duration of this scan.}\label{fig:figure1}
        \end{figure*}

 \begin{figure*}[ht]
            \centering\includegraphics[width=0.95\textwidth]{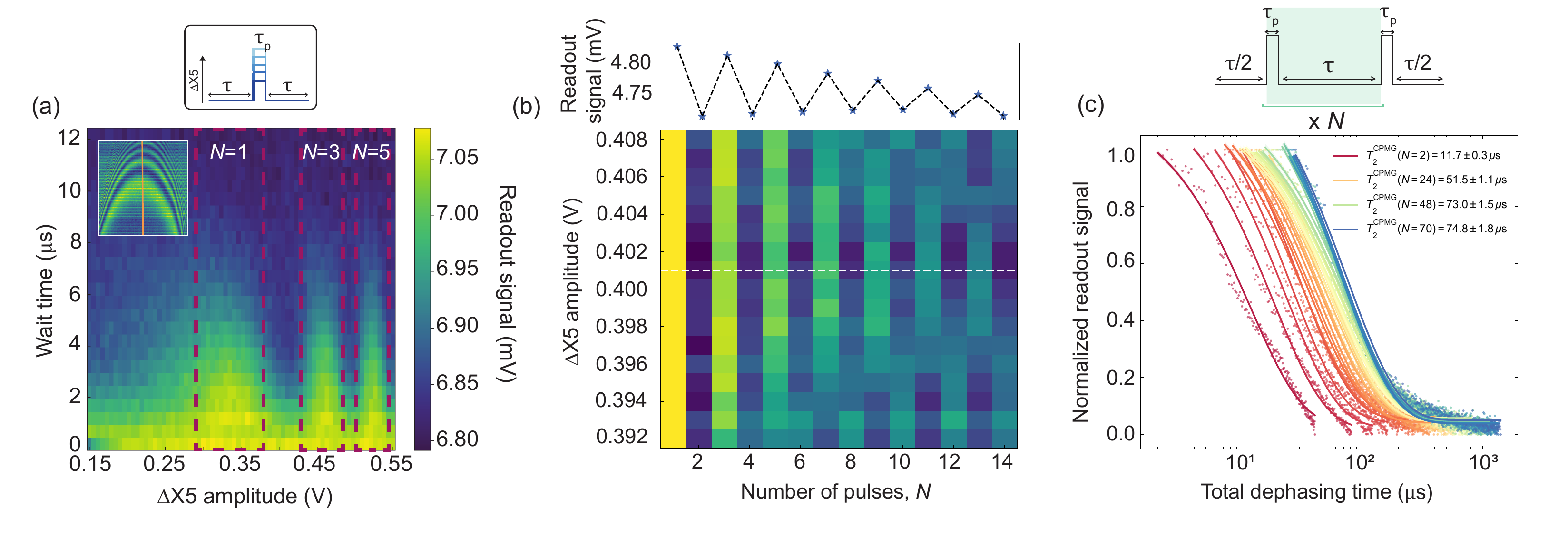} 
            \caption{\textbf{Characterization of the double quantum dot device.} (a) 2D sweep of exchange pulse $\Delta$X5 amplitude against wait time in the dephasing (1,1) region, with the $\Delta$P6, $\Delta$P5 values set to 0V. The $\Delta$X5 sweep axis therefore represents a linecut through the exchange fingerprint shown in the inset. For amplitudes that correspond to an odd number of $\pi$-pulses, a refocusing effect occurs, which results in a Hahn echo-like sequence that extends the singlet coherence time. These refocusing amplitudes are highlighted in the red dashed boxes. For amplitudes that correspond to an even number of $\pi$-pulses, the exchange pulse has no refocusing effect, effectively turning the experiment into a Ramsey-style sequence, with a correspondingly lower coherence. (b) Calibration of the exchange $\pi$-pulse by repeating multiple $\pi$-pulses, while sweeping the $\pi$-pulse amplitude. Results are normalized across the amplitude rows. Note that a different exchange pulse time was used compared to the measurement shown in (a), resulting in a $\pi$-pulse occurring at a different $\Delta$X5 amplitude. From this sequence an optimal $\Delta$X5 amplitude is chosen, as shown by the white dashed line. The upper panel shows a $\pi$-pulse train using the optimal $\Delta$X5 amplitude extracted from (b). (c) CPMG sequence up to \textit{N}=70 pulses, using the calibrated $\pi$ exchange pulses. A maximum $T_2^{\text{CPMG}}$ time of 74.8 $\pm$ 1.8 \unit{\micro\second} was extracted. Upper panel shows the pulse sequence used to obtain the CPMG data. The total dephasing time represents the total wait time across the entire CPMG sequence, i.e. $N \times \tau$. Each curve was fit with an exponential decay function to extract the decay time and normalized between 0 and 1 before plotting. The decay-time uncertainty is estimated as the square root of the corresponding diagonal element of the fit covariance matrix, providing an approximate one-standard-deviation uncertainty under the assumptions of the least-squares model. }\label{fig:figure2}
\end{figure*}

In this work, we use a double quantum dot (DQD) Si/SiGe single-layer etch-defined gate electrode (SLEDGE) device fabricated at HRL Laboratories \cite{ha2021flexible, weinstein2023universal}. Figure~\ref{fig:figure1}(a) shows a false-colored scanning electron microscope (SEM) image of a device nominally identical to the device used in this work. Although the device comprises a linear array of six quantum dots, only the dot plunger gates P4, P5, and P6 were connected to external control lines. Throughout this work, we focus exclusively on the double quantum dot defined beneath gates P5 and P6, while gate P4 is held at 0~V, to deplete electrons from beneath this gate. The exchange interaction between the P5 and P6 dots is controlled using the X5 barrier gate. Electrons are loaded from an ohmic reservoir (not shown) via bath gate B3, with the tunnel coupling to the bath tuned by the T2 barrier gate. The outer field gates (OFG, pink) and inner field gate (IFG, yellow) are DC biased in depletion mode to provide lateral electrostatic confinement of the quantum dots.

Spin-state readout is performed using Pauli spin blockade (PSB) between the P5 and P6 quantum dots \cite{ono2002current}. The resulting charge state is detected by a nearby dot charge sensor (DCS), formed by a large quantum dot beneath the M2 plunger gate along with two barrier gates, Z3 and Z4. A 400~kHz excitation is applied to the source ohmic contact, and the corresponding drain signal is demodulated at the same frequency to measure the DCS conductance \cite{blumoff2022fast}.

The device is operated at zero externally applied magnetic field, with the qubit encoded in the singlet and triplet spin states of the electrons confined beneath gates P5 and P6. We therefore work in the $\{\ket{S},\ket{T_0}\}$ basis. Although no magnetic field is intentionally applied, the unshielded ambient geomagnetic field, $B_{\mathrm{amb}}\approx58~\mu\mathrm{T}$, provides a residual quantization axis and a Zeeman separation $g\mu_{\text{B}}B_{\mathrm{amb}}/h\approx1.6~\mathrm{MHz}$ between $T_0$ and $T_{\pm}$. Since this substantially exceeds the maximum nuclear-gradient we expect in this system (see Section \ref{sec:3}), mixing with $T_{\pm}$ is expected to be suppressed, justifying an effective $\{\ket{S},\ket{T_0}\}$ description.

Qubit control is achieved by driving exchange oscillations between these two quantum dots. Conventionally, exchange oscillations in double quantum dots are performed by combining exchange control with an externally applied magnetic field gradient $\Delta B_z$, of magnitude approximately 10 mT \cite{struck2020low, zajac2018resonantly}, across the dots. This magnetic field gradient provides a second control axis orthogonal to the exchange interaction $J$ enabling universal single-qubit control. In this case, the Hamiltonian of the double quantum dot system is

\begin{equation}
H
=
\frac{\hbar}{2}
\left(
\omega_{B_z}\sigma_z
-
\omega_J\sigma_x
\right),
\end{equation}

where
\begin{eqnarray}
\label{Ham}
\omega_{B_z} &=& \frac{g\mu_{\text{B}}\Delta B_z}{\hbar}, \label{eq:omega_b}\\
\omega_J &=& \frac{J}{\hbar}.
\end{eqnarray}

We perform exchange oscillations between quantum dots at zero applied field by utilizing the intrinsic $\Delta B_{z}$ introduced by the nuclear spin bath surrounding the quantum dots \cite{petta2005coherent, maune2012coherent, berritta2024real, laird2006effect}. An intrinsic magnetic-field gradient $\Delta B_z$ exists between the quantum dots even in the absence of an externally applied magnetic field. This gradient primarily arises from the residual 800 ppm concentration of spinful $^{29}$Si nuclei in the isotopically enriched silicon host and $^{73}$Ge nuclei in the barriers surrounding the silicon well, which couple to the electron spins through the hyperfine interaction \cite{struck2020low, eng2015isotopically, witzel2012nuclear, kerckhoff2021magnetic}. Because their centers are separated by approximately $100~\mathrm{nm}$, the two dots sample distinct local nuclear-spin environments, resulting in a difference between the effective Overhauser fields experienced by the electrons. For dots with a similar separation, the interdot $g$-factor difference of $\Delta g \approx 6\times10^{-4}$ reported in similar devices \cite{connors2022charge}, would produce a frequency difference of only approximately $ 500~\mathrm{Hz}$ in Earth’s ambient magnetic field. This is significantly smaller than the effect attributed to the nuclear spins in our measurements, indicating that the interdot magnetic-field gradient in our system is dominated by the nuclear-spin bath rather than by the difference in the electron $g$-factors.

Figure~\ref{fig:figure1}(b) illustrates the pulse sequence used to drive exchange oscillations in the double quantum dot. We use a Bloch-sphere convention rotated relative to the standard singlet-triplet representation, with $\Delta B_z$ aligned along the z-axis, so that phase accumulation, and the resulting dephasing, during free-evolution is represented naturally as precession about z, as described below.

A singlet state is first initialized beneath gates P5 and P6 by waiting at the (0,1)-(0,2) charge transition, where the charge occupancy is denoted by $(N_{\mathrm{P5}}, N_{\mathrm{P6}})$, for a duration longer than the singlet relaxation time (see Appendix Section~\ref{sec:init} for details on singlet initialization optimization). The gate voltages on P5 and P6 are then ramped across the (0,2)-(1,1) interdot transition before pulsing deep into the (1,1) charge configuration. At this operating point, the overlap of the two electron wavefunctions is reduced, causing the exchange interaction $J$ to be strongly suppressed. 

After pulsing into the (1,1) charge configuration, the spin state is allowed to evolve freely for a time $\tau$. During this free-evolution period, the exchange coupling is much smaller than the magnetic-field gradient, $J \ll g\mu_{\text{B}} \Delta B_z$, and therefore the spin state precesses about the intrinsic magnetic field gradient $\Delta B_z$ on the singlet-triplet Bloch sphere, as illustrated in Fig.~\ref{fig:figure1}(b).

Following this free-evolution period, simultaneous voltage pulses are applied to the X5, P5, and P6 gates to rapidly increase the exchange interaction strength such that the exchange interaction becomes much larger than the magnetic-field gradient, $J \gg g\mu_{\text{B}}\Delta B_z$. These voltage pulses are superimposed on static DC offsets applied to each gate. The spin state then evolves for an exchange pulse duration $\tau_{\text{p}}$, precessing about an axis primarily pointing along the exchange axis of the Bloch sphere (the +x axis in our Bloch sphere notation). After the exchange pulse, the gate voltages are returned to the original (1,1) operating point, and the spin state undergoes a second period of free-evolution for a further time $\tau$.

This pulse sequence is analogous to a Hahn-echo experiment, with the exchange pulse acting as the refocusing pulse for certain exchange pulse parameters. When the exchange pulse produces a $(2n+1) \times \pi$-rotation about the exchange axis, where $n$ is an integer, the dephasing accumulated during the first free-evolution period is reversed during the second, resulting in a refocusing of the spin state. By sweeping either the amplitudes of the X5, P5, and P6 voltage pulses, and hence the magnitude of $J$, or the pulse duration $\tau_{\mathrm{p}}$, the exchange rotation is tuned through successive odd and even multiples of $\pi$, alternately refocusing and not refocusing the spin state and thereby producing the observed exchange oscillations.

Importantly, this refocusing mechanism remains effective because of the slow fluctuations in $\Delta B_z$ arising from nuclear spin dynamics. Although $\Delta B_z$ evolves over long timescales due to nuclear spin flip-flops, it can be treated as quasi-static during a single execution of the pulse sequence \cite{petta2005coherent}. Under this quasi-static approximation, the same value of $\Delta B_z$ is experienced across both free-evolution periods, allowing the accumulated phase to be refocused by the $\pi$ exchange pulse \cite{kerckhoff2021magnetic}. A more detailed theoretical discussion of this technique can be found in Appendix Section~\ref{sec:a_derivation}. 

We use the pulse sequence described above to measure exchange oscillations between the quantum dots beneath gates P5 and P6. Figure~\ref{fig:figure1}(c) shows a nonequilibrium $J_{\mathrm{X5}}$ map acquired for a fixed $\Delta$X5 = 0.55 V, while sweeping the $\Delta$P5 and $\Delta$P6 gate voltages. This measurement is used to identify the detuning axis (shown by the red arrow) along which P5 and P6 are swept to obtain the exchange fingerprint measurements shown in Fig.~\ref{fig:figure1}(d-f).

Although the exchange-oscillation pattern is, in principle, agnostic to the exact wait time in the dephasing region, its visibility is governed by the phase dispersion accumulated under $\Delta B_z$ during each free-evolution period, as described in Appendix Section~\ref{sec:a_derivation}. To obtain appreciable visibility, $\tau$ should be a significant fraction of the characteristic gradient-evolution timescale $\tau_{\Delta B_z}\equiv 1/\sigma_{B_z}$, where $\sigma_{B_z}$ is the standard deviation of the angular-frequency distribution $\omega_{B_z}=g\mu_{\mathrm{B}}\Delta B_z/\hbar$ arising from quasistatic, shot-to-shot fluctuations in the interdot nuclear magnetic-field gradient. When $\tau$ approaches this timescale the spin state evolves appreciably away from the initialized singlet state. Maximum contrast is obtained when $\tau$ is of order $T_2^*$ or longer ($\tau\gtrsim T_2^*$), such that the ensemble-averaged state becomes approximately fully dephased during free evolution. Sweeping the exchange-pulse area can then vary the singlet-return probability between the fully mixed value, $P_{\mathrm{S}}\approx1/2$, and the refocused singlet value, $P_{\mathrm{S}}\approx1$. The free-evolution time should nevertheless remain shorter than the Hahn-echo coherence time ($\tau<T_2^{\mathrm{Hahn}}$), so that the quasi-static nuclear-field gradient can still be effectively refocused by the exchange pulse.

This behavior is illustrated in Fig.~\ref{fig:figure1}(d-f). For a short free-evolution time of $\tau$ = 500 ns [Fig.~\ref{fig:figure1}(d)], which is below the measured $T_2^* = 1.17 \pm 0.02$ \unit{\micro\second}, the spin state remains predominantly in the singlet state during the free-evolution, resulting in low contrast exchange oscillations. Increasing the free-evolution time to $\tau = 2~$\unit{\micro\second} [Fig.~\ref{fig:figure1}(e)], which is beyond $T_2^*$, allows for more substantial evolution under $\Delta B_z$, producing higher-contrast oscillations. Finally, for $\tau = 6~$\unit{\micro\second} [Fig.~\ref{fig:figure1}(f)], the free-evolution time begins to approach the Hahn-echo coherence limit. On this timescale, the assumption of a quasi-static nuclear field begins to break down, resulting in dephasing during the free-evolution period that cannot be refocused by the exchange pulse and hence the observation of dephased oscillations. AC electric noise could also contribute to this dephasing. 

Figure~\ref{fig:figure1}(g) shows the exchange oscillations as a function of the X5 pulse amplitude and duration. By extracting the oscillation frequency at each $\Delta$X5 amplitude, we determine an exchange tunability of $5.2~\mathrm{decades/V}$. We also extract the exchange-oscillation quality factor (Q-factor), defined here as the number of oscillation periods completed within the characteristic decay time. Over the voltage range studied, the quality factor increases with the applied X5 amplitude (see Appendix Section~\ref{sec:eo}).

 \begin{figure*}[ht]
            \centering\includegraphics[width=1\textwidth]{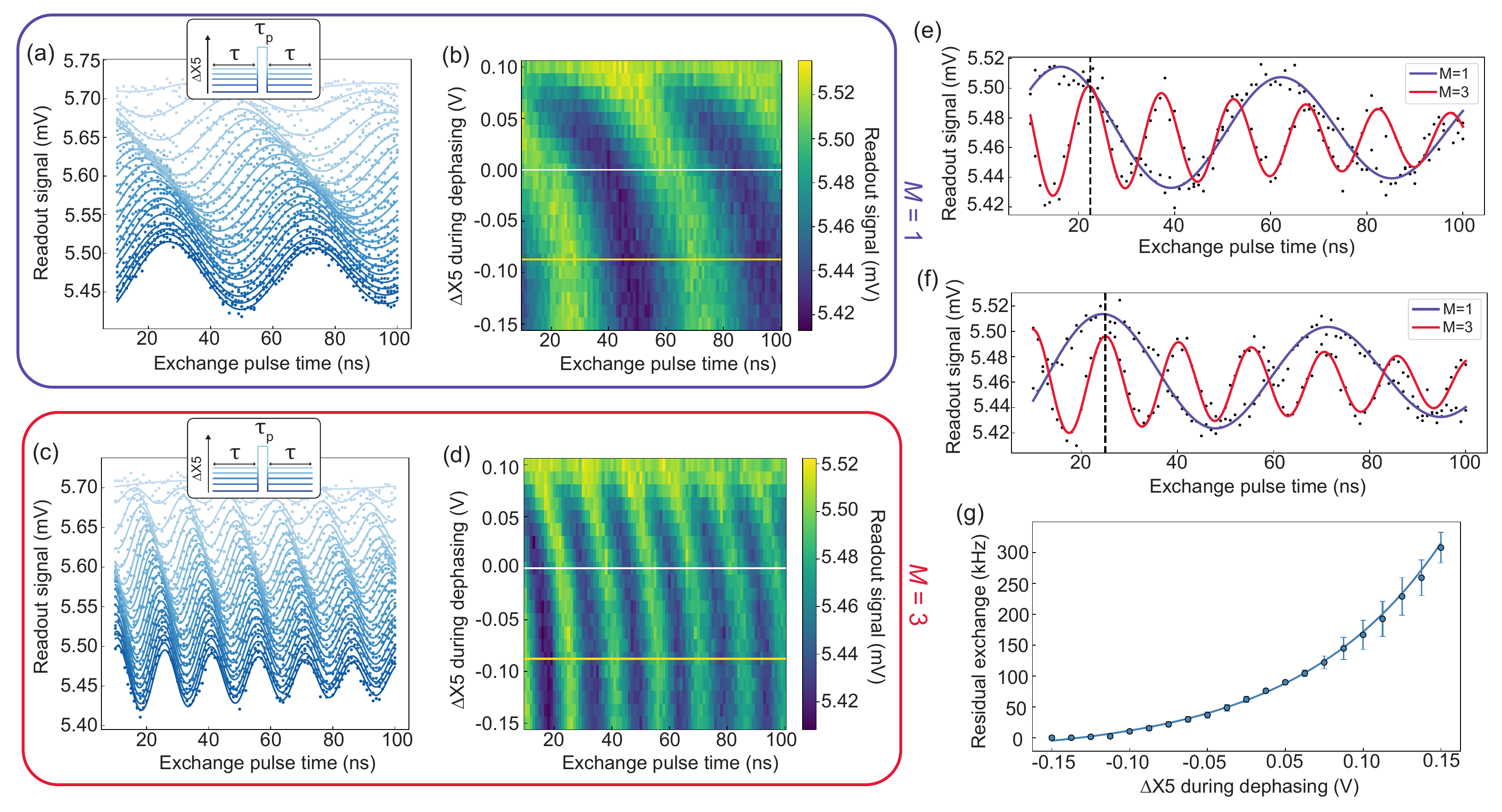} 
            \caption{\textbf{Probing low values of residual exchange.} (a) A sequence consisting of waiting in the dephasing region in (1,1) for 2 \unit{\micro\second}, before applying a variable length pulse on the X5 exchange gate, followed by a further wait time in the (1,1) dephasing region for 2 \unit{\micro\second}. Each vertically offset curve in the waterfall plot represents a different $\Delta$X5 amplitude applied during the two dephasing wait times. The $\Delta$X5 amplitude during the exchange pulse was fixed. It should be noted that all $\Delta$X5 amplitudes discussed are applied on top of a constant X5 DC offset that is applied throughout the experiment and was determined when tuning up the device for PSB readout. Dark blue curves represent low $\Delta$X5 amplitudes applied during dephasing, where the residual exchange is expected to be suppressed, and light blue curves higher $\Delta$X5 amplitudes, for which we expect to observe higher residual exchange. Pulse sequence inset denotes that we are sweeping the X5 amplitude during the free-evolution periods in these scans. This plot is also shown for clarity in a 2D color plot in (b). (c, d) The same measurement performed in (a) and (b), but with three rather than one refocusing exchange pulses applied one after the other, with 20 ns in between each pulse. (e) 1D linecuts through (b) and (d) for $\Delta$X5=0V (upper white lines). Purple and red data correspond to $M$=1 and $M$=3 pulses, respectively. (f) 1D linecuts through (b) and (d) for $\Delta$X5=$-$0.09V (lower yellow lines). Purple and red data correspond to $M$=1 and $M$=3 pulses, respectively. (g) Estimated residual exchange values extracted from a joint Hamiltonian fit to each row in (b) and (d) (see Appendix Section \ref{sec:simulation}). } \label{fig:figure3}
        \end{figure*}

 \begin{figure*}[ht]
         \centering\includegraphics[width=1\textwidth]{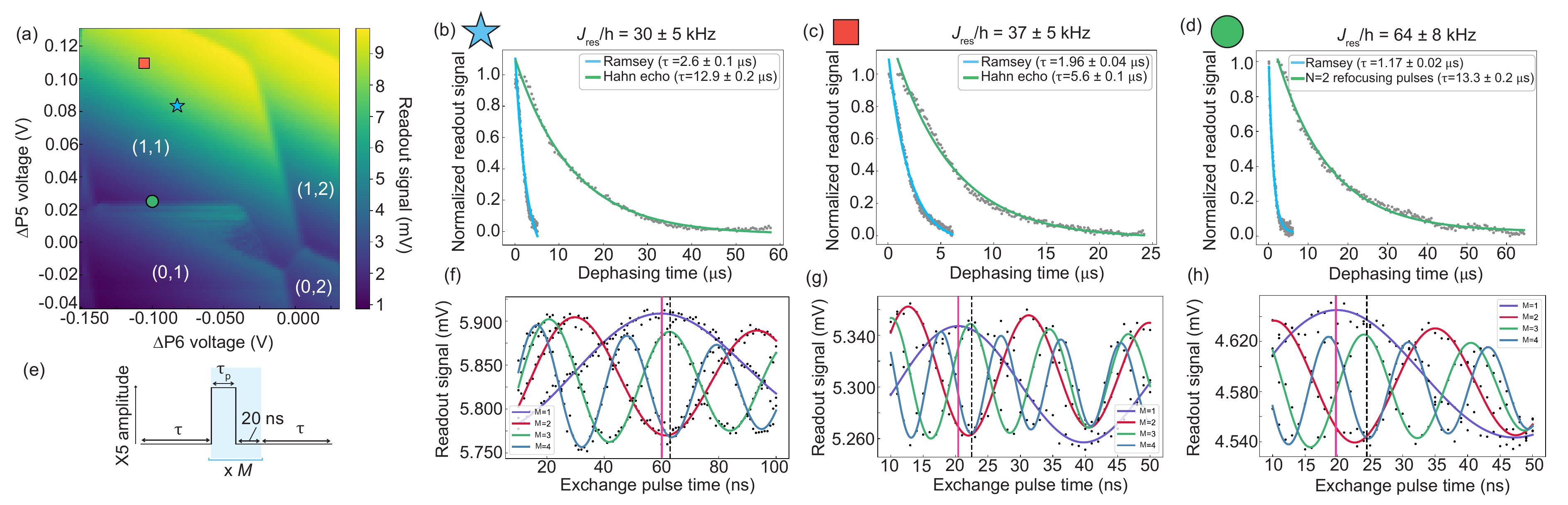} 
            \caption{\textbf{Characterizing the (1,1) charge cell.} (a) Charge stability map around the (1,1)-(0,2) interdot transition. The red square, blue star and green circle represent points in the (1,1) charge cell used for the dephasing point in the exchange pulses sequences. (b) Ramsey (blue curve) and Hahn echo measurement (green curve) performed with the dephasing point at the blue star point shown in (a). (c) Ramsey (blue curve) and Hahn echo measurement (green curve) performed with the dephasing point at the red square point shown in (a). (d) Ramsey (blue curve) and CPMG measurement with two refocusing pulses (green curve) performed with the dephasing point at the green circle point shown in (a). (e) Pulse sequence used to obtain the plots in (f-h). (f,g,h) Sweeping the time of the exchange $\Delta$X5 pulse for \textit{M}=1 (purple line), \textit{M}=2 (red line), \textit{M}=3 (green line) and \textit{M}=4 (blue line) exchange pulses with the dephasing point at the (f) blue star point, (g) red square point, and (h), green circle point in (a). The vertical pink line in each plot denotes the maxima of the $M=1$ curve, while the vertical dashed black line denotes the maxima of the $M=3$ curve. The error bars in all coherence measurements represent one standard deviation uncertainty of the decay time from the fit.}\label{fig:figure4}
\end{figure*}

\section{\label{sec:2} Pulse and dephasing characterization}

Utilizing the intrinsic magnetic field gradient $\Delta B_z$ generated by the local nuclear spin bath to observe exchange oscillations provides a simple means of characterizing the coherence of the two-spin system and, consequently, the underlying noise environment. Figure~\ref{fig:figure2}(a) measures the coherence time along a one-dimensional line cut through the exchange fingerprint scan (shown inset), obtained by sweeping the $\Delta$X5 pulse amplitude while holding the P5 and P6 gate voltages fixed at the center of the fingerprint. We do not perform gate virtualization for these scans and therefore pulsing on gate X5 will also affect the potentials of the P5 and P6 quantum dots. The exchange pulse duration $\tau_{\text{p}}$ was fixed at 100 ns.

At specific $\Delta$X5 pulse amplitudes, the enhanced singlet return probability persists over longer free-evolution times, indicating Hahn-echo refocusing. These operating points, highlighted by the red dashed boxes in Fig.~\ref{fig:figure2}(a), occur when the exchange pulse implements an odd multiple of a $\pi$-rotation and thereby reverses the phase accumulated during the two free-evolution periods. The rotation angle is controlled by the $\Delta$X5 pulse amplitude, which tunes the strength of the exchange interaction and hence the rotation rate about the exchange axis of the singlet-triplet Bloch sphere. 

In contrast, exchange pulses corresponding to even multiples of a $\pi$-rotation do not refocus the accumulated phase. These operating points are therefore equivalent to Ramsey-type measurements, with correspondingly shorter singlet coherence observed. 

We extend this approach beyond Ramsey and Hahn-echo measurements by implementing CPMG sequences containing up to $N$=70 refocusing pulses. Accurate calibration of the exchange $\pi$-pulses is essential to maintain high-fidelity refocusing over such long pulse trains. To achieve this, we employ the repeated-pulse calibration measurement shown in Fig.~\ref{fig:figure2}(b). An initial estimate of the $\Delta$X5 $\pi$-pulse amplitude and duration is obtained from the exchange fingerprint measurement. Using these parameters, we apply a sequence of 14 nominal-$\pi$ exchange pulses, separated by 20 ns free-evolution intervals, while sweeping the $\Delta$X5 pulse amplitude about its initial estimate. Deviations from the ideal $\pi$-pulse condition accumulate over the pulse train, making the measurement sensitive to small calibration errors. The optimal $\Delta$X5 pulse amplitude is indicated by the white dashed line in Fig.~\ref{fig:figure2}(b). The corresponding repeated-pulse measurement acquired at this calibrated amplitude is shown in the upper panel.

Figure~\ref{fig:figure2}(c) shows the CPMG measurements for \textit{N}=2-70 refocusing pulses, using the calibrated $\pi$-pulse from Fig.~\ref{fig:figure2}(b). The singlet coherence time was extended up to $T_2^{\text{CPMG}}= 74.8 \pm 1.8~$\unit{\micro\second} for \textit{N}=70 refocusing pulses. Note that this CPMG sequence was carried out with the idle position at the circle point in the (1,1) cell, discussed in Section \ref{sec:3}. A slow drift in the baseline readout signal at long dephasing times distorted the CPMG decay envelopes, preventing a reliable quantitative interpretation of their functional form. The origin of this drift remains unclear, although residual exchange at the idle point and gate voltage drift may contribute.

\section{\label{sec:3} Probing residual exchange}
The phase of the exchange oscillations provides a highly sensitive probe of the exchange interaction present during idling in the (1,1) charge configuration, which we refer to as residual exchange \cite{watson2018programmable, hamonic2026single}. Any finite exchange interaction during these intervals contributes an additional rotation about the exchange axis of the Bloch sphere, resulting in an accumulated phase that shifts the position of the exchange oscillations. Detecting and accounting for these low values of residual exchange while idling in the (1,1) cell is important for implementing high-fidelity operations. Of particular relevance to the CPMG sequences shown in the previous section, residual exchange that is barely visible in a single-echo measurement produces a coherent error that grows under repeated refocusing and limits the achievable dynamical-decoupling depth.

To investigate this effect, we intentionally modify the residual exchange by applying a voltage pulse to the X5 barrier gate throughout both free-evolution periods while sweeping its amplitude. It should be noted that, as for the exchange oscillations, this $\Delta$X5 voltage pulse is applied on top of a constant DC offset that is applied throughout the experiment. For sufficiently negative $\Delta$X5 pulse voltages, the tunnel barrier between the dots is increased, suppressing the residual exchange interaction. As the $\Delta$X5 pulse voltage is increased, the tunnel barrier is lowered, and the residual exchange is expected to increase.

Figures~\ref{fig:figure3}(a,b) show measurements performed using \textit{M}=1 pulses, where \textit{M} is the number of exchange pulses applied between the two free-evolution periods. The spin state first evolves freely in the (1,1) charge configuration for $\tau = 2$~\unit{\micro\second} before a single exchange pulse of variable duration is applied, followed by a second 2~\unit{\micro\second} free-evolution period. During both free-evolution intervals, the additional $\Delta$X5 pulse described above is applied, with its amplitude swept to control the residual exchange. The data are presented as a vertically offset waterfall plot in Fig.~\ref{fig:figure3}(a) and as a two-dimensional color map in Fig.~\ref{fig:figure3}(b). Figures~\ref{fig:figure3}(c,d) show the equivalent measurements performed using \textit{M}=3 refocusing pulses, with each pulse separated by 20~ns.

To investigate the effect of the residual exchange, Fig.~\ref{fig:figure3}(e) compares one-dimensional line cuts extracted from the upper white lines in Figs.~\ref{fig:figure3}(b) and (d), corresponding to $\Delta$X5 = 0 V during the free-evolution periods. Under these conditions, a finite residual exchange remains in the (1,1) charge configuration. The additional phase accumulated about the exchange axis shifts the exchange oscillations, such that the maxima of the \textit{M}=1 (purple) and \textit{M}=3 (red) measurements occur at different exchange pulse durations.

In contrast, Fig.~\ref{fig:figure3}(f) compares line cuts extracted from the lower yellow lines in Figs.~\ref{fig:figure3}(b) and (d), corresponding to an $\Delta$X5 pulse voltage of $\Delta$X5= $-$90 mV during the free-evolution periods. At this operating point, the increased tunnel barrier suppresses the residual exchange interaction, reducing the unwanted phase accumulation about the exchange axis. Consequently, the \textit{M}=1 and \textit{M}=3 oscillation maxima coincide, as expected in the absence of significant residual exchange.

We estimate the residual exchange at each $\Delta$X5 amplitude by jointly fitting the \textit{M}=1 and \textit{M}=3 refocusing measurements to a Hamiltonian time-evolution simulation (see Appendix Section~\ref{sec:simulation}). The joint fitting function also extracts a shared value of $\sigma_{B_z}$, which represents the width of the Gaussian distribution of $\omega_{B_z} = g \mu_\text{B}\Delta B_z/\hbar$. From the joint fit we obtain $\sigma_{B_z}/2\pi = $140 $\pm$ 6 kHz. This $\sigma_{B_z}/2\pi$ value is of the same order as, although somewhat larger than, magnetic-gradient widths previously reported for Si/SiGe devices containing 800 ppm residual $^{29}\mathrm{Si}$ \cite{kerckhoff2021magnetic, eng2015isotopically}. Figure~\ref{fig:figure3}(g) shows the extracted residual exchange frequency as a function of the $\Delta$X5 amplitude applied during free-evolution, revealing the expected exponential dependence on voltage. From this, we estimate a residual exchange frequency of approximately 100~kHz at a $\Delta$X5 amplitude of 0~V for this operating point in the $(1,1)$ charge cell.

Having established that exchange oscillations provide sensitive probes of the residual exchange during the free-evolution period, we next use this technique to characterize the (1,1) charge configuration used during idling. Specifically, we investigate how the coherence time and residual exchange interaction vary as the free-evolution point is moved throughout the (1,1) charge cell.

Figure~\ref{fig:figure4}(a) shows the charge stability diagram centerd on the (1,1) charge configuration, with three representative operating points highlighted. At each location, we first characterize the spin coherence using Ramsey and refocusing measurements. Figures~\ref{fig:figure4}(b) and (c) compare Ramsey (blue) and Hahn-echo (green) measurements acquired at the blue star and red square idling points, respectively, while Fig.~\ref{fig:figure4}(d) shows Ramsey (blue) and \textit{N}=2 CPMG (green) measurements acquired at the green circle operating point. 

We next use the phase of the exchange oscillations to quantify the residual exchange at each idling point. Figure~\ref{fig:figure4}(e) shows the pulse sequence used to carry out the measurements in (f-h). Figures~\ref{fig:figure4}(f-h) show exchange oscillations measured with \textit{M}=1, 2, 3 and 4 exchange pulses, without applying an additional $\Delta$X5 pulse during the free-evolution periods. In the absence of residual exchange, the oscillation maxima for \textit{M}=1 and \textit{M}=3, together with the minima for \textit{M}=2 and \textit{M}=4, are expected to coincide, as each sequence accumulates the same exchange phase during free-evolution. This behavior is observed only for the blue star operating point, indicating lower residual exchange near the center of the (1,1) charge configuration. Note that a different DC X5 offset was used in this measurement compared to the measurement shown in Fig.~\ref{fig:figure3}, explaining the difference in residual exchange observed at $\Delta$X5=0V between these measurements. In contrast, the increasing misalignment of the oscillation extrema, as shown by the distance between the vertical pink and black dashed lines in Figs.~\ref{fig:figure4}(g,h), at the red square and green circle operating points, demonstrates that the residual exchange increases towards the edges of the (1,1) charge cell, as expected \cite{reed2016reduced, martins2016noise}.

We again use a joint time evolution simulation fit for \textit{M}=1,2,3,4 (see Appendix Section~\ref{sec:simulation}) to extract an estimated residual exchange value for each operating position. For the joint fit, we use the value $\sigma_{B_z}/2\pi=140\pm6~\mathrm{kHz}$ extracted from the fits in Fig.~\ref{fig:figure3}(g). We test the consistency of this value by calculating the corresponding dephasing time, assuming that dephasing is dominated by quasi-static nuclear-spin fluctuations with a Gaussian frequency distribution. Neglecting energy relaxation over the relevant timescale, such that $T_2^* \simeq T_\phi$, and assuming a quasi-static noise profile, we model the Ramsey decay as $e^{-(t/T_2^*)^2}$. The dephasing time is then related to the standard deviation
of this distribution by $T_2^*=\sqrt{2}/\sigma_{B_z}$. This gives $T_2^*=1.6$ \unit{\micro\second}, consistent with the
range of values measured in Figs.~\ref{fig:figure4}(b)-(d).

Using this $\sigma_{B_z}$ estimate, we extract residual exchange values of $J_{\text{res}}/h$ = 30 $\pm$ 5 kHz, 37 $\pm$ 5 kHz and 64 $\pm$ 8 kHz for the star, square and circle idling points, respectively, using the joint $M$=1-4 fitting function. This supports our qualitative assessment from the misalignment of the \textit{M}=1-4 curves that the circle operating point possesses the highest residual exchange and is consistent with this position's proximity to the edge of the (1,1) cell.

The observed anti-correlation between $T_2^*$ time of the three operating points and the estimated residual exchange suggests that charge noise may have a non-negligible contribution to dephasing across the (1,1) charge cell.

\section{\label{sec:4} Singlet survival probability vs exchange frequency}

 \begin{figure}[h]
            \centering\includegraphics[width=0.45\textwidth]{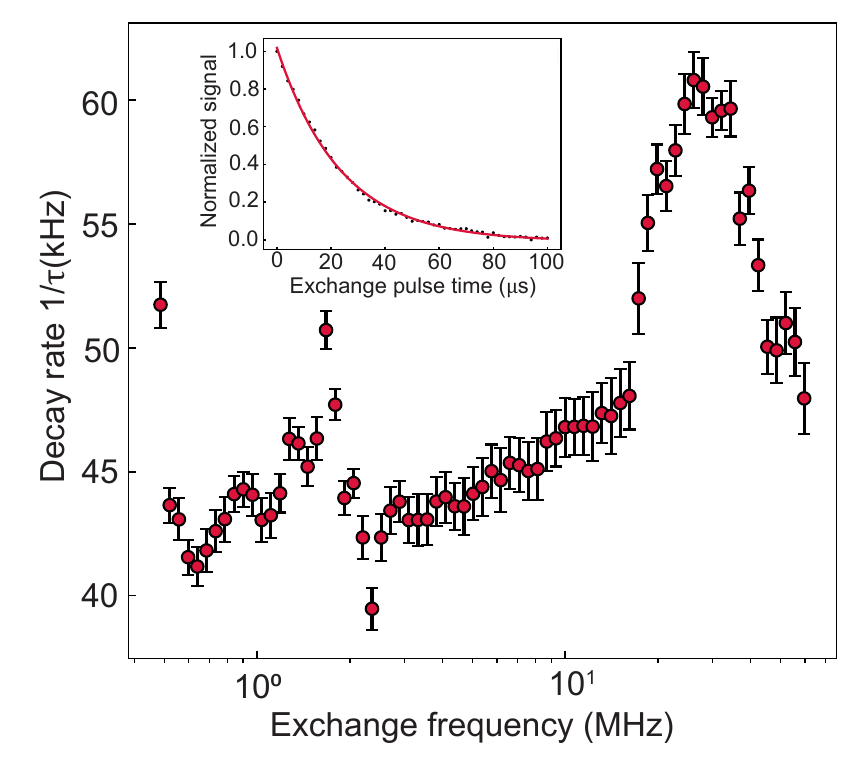} 
            \caption{\textbf{Singlet dephasing rate against exchange frequency.} Plot of the singlet decay rate against exchange interaction frequency. This experiment involves initializing a singlet state between the P5 and P6 dots before turning on the exchange interaction for a variable length of time. The decay time is extracted by fitting the decay of the singlet as a function of exchange pulse time. An example such decay curve is shown in the inset. The exchange pulse amplitude is then swept, with the decay time extracted for each amplitude. This amplitude is then converted to an equivalent exchange interaction strength. Error bars are obtained by the standard deviation of a bootstrapped distribution of the decay rates.}\label{fig:figure5}
        \end{figure}

To characterize the noise and relaxation processes that limit exchange-based control in the device, we measure the decay of the singlet-return signal during an exchange pulse and extract an effective exchange-dependent singlet decay time as a function of exchange frequency \cite{dial2013charge}. 

To obtain this decay spectrum, we first initialize the double quantum dot in the singlet state before transferring the system to the (1,1) charge configuration. We then pulse gates X5, P5, and P6 simultaneously to turn on the exchange interaction.  Unlike earlier sequences, no free-evolution takes place between preparation in (1,1) and the exchange pulse and thus no oscillations are observed. For each exchange-pulse amplitude, we sweep the pulse duration and fit the resulting averaged singlet-return signal to an exponential decay. From the fitted decay time $\tau_{\text{S}}$, we define the effective singlet decay rate during the exchange pulse as, $\Gamma_{\text{S}} = 1/\tau_{\text{S}}$. This measurement is repeated over a range of $\Delta$X5 pulse amplitudes to characterize the dependence of the decay rate on the exchange interaction.

We independently calibrate the relationship between $\Delta$X5 pulse amplitude and exchange frequency using the method shown in Appendix Section \ref{sec:eo}, for the specific bias point investigated. This calibration is then used to convert $\Delta$X5 pulse amplitudes within the measured range into exchange frequencies.

Figure~\ref{fig:figure5} shows the extracted singlet decay rate as a function of exchange frequency. The decay rate exhibits a broad maximum between approximately 20 and 40 MHz exchange frequency, corresponding to a reduction in the singlet decay time. Two plausible explanations for this are: 1) an enhanced relaxation or leakage caused by hybridization with an additional state, and 2) a broad enhancement of longitudinal dephasing noise at the exchange splitting \cite{yang2013spin, huang2014spin, raith2012theory, kornich2018phonon}. These mechanisms are not mutually exclusive, as state hybridization may increase the coupling of the system to electrical or phononic fluctuations.

In the first scenario, tuning the barrier modifies the charge, orbital, or valley composition of the singlet-like eigenstate. Hybridization with an additional state can increase the matrix element coupling the singlet state to environmental fluctuations or open a leakage channel outside the measured subspace. The resulting increase in the transition rate can produce an observed reduction in lifetime. This hypothesis could be tested in future work by mapping the decay rate against a second control parameter, such as gate detuning or magnetic field.

Alternatively, magnetic-gradient noise or electrical noise converted into longitudinal coupling through spin-orbit or excited-state mixing could drive transitions out of the singlet-like state over this frequency range. This possibility could be tested by reaching the same value of $J$ using different gate-voltage configurations. A feature that remains fixed in exchange frequency would support an environmental noise-spectrum origin. Injecting calibrated noise could additionally be used to measure the frequency-dependent susceptibility of the system and determine whether the observed decay is consistent with longitudinal-noise-driven transitions.

\section{\label{sec:5} Conclusion}

We have demonstrated a method for characterizing exchange-based semiconductor qubits at zero applied magnetic field using time-averaged exchange oscillations driven by the intrinsic gradient produced by residual nuclear spins. Rather than requiring calibrated control about multiple exchange axes, the technique uses simple exchange-pulse sequences and singlet-return measurements to isolate different contributions to the underlying device dynamics. We demonstrate this approach through three complementary measurements in a Si/SiGe double quantum dot. First, CPMG sequences containing up to \textit{N}=70 refocusing pulses extend the singlet coherence from $T_2^*=1.17\pm0.02~$\unit{\micro\second} to $T_2^{\mathrm{CPMG}}=74.8\pm1.8~$\unit{\micro\second}. Second, analysis of the phase response for sequences containing \textit{M}=1-4 refocusing pulses resolves residual exchange coupling in the tens-of-kilohertz regime during nominally idle evolution and enables it to be mapped across the (1,1) charge cell. Finally, measurements of the singlet-return decay under constant exchange reveal reproducible structure in the decay rate as a function of exchange frequency, including a broad enhancement between approximately 20 and 40 MHz of exchange coupling. Complementary measurements could determine whether this feature arises from longitudinal environmental noise at the exchange splitting or from relaxation and leakage associated with hybridization with additional states.

More broadly, although demonstrated here in a double quantum dot using singlet and triplet spin states, the method provides a simple characterization primitive for exchange-only and related zero-applied field encodings constructed from singlet-like and triplet-like logical states. Device noise characteristics, idle interactions, and decay mechanisms can therefore be assessed independently before, or alongside, calibration of the complete encoded-qubit control space. Its operation without an intentionally applied magnetic field, micromagnet, or dynamically prepared nuclear gradient makes the approach particularly well suited to semiconductor qubits intended for integration with field-sensitive superconducting resonators and qubits. In a superconducting-semiconductor hybrid system, magnetic shielding may reduce the ambient field below the characteristic hyperfine field considered here. In this regime, mixing with $T_{\pm}$ states would produce qualitatively different dynamics and require an expanded treatment beyond the $S$-$T_0$ subspace, but the underlying characterization approach would remain applicable.

\begin{acknowledgments}

The authors are grateful to Joe Kerckhoff, Edwin Acuna,  Marc Dvorak, Arne Laucht and Fabrizio Berritta for helpful discussions and to Ilan Rosen and Mathis Moes for help with infrastructure development. The device was fabricated and provided by HRL Laboratories, LLC, as part of the Qubits for Computing Foundry program, which is a collaborative effort under the Army Research Office (ARO) and the Laboratory for Physical Sciences (LPS) Qubit Collaboratory. This material is based upon work supported by the National Science Foundation under Grant No. 2427120. HHK is supported by Korea Foundation for Advanced Studies.
\end{acknowledgments}

\section*{Data availability}
The data that support the findings of this article are publicly available at \cite{stemp2026exchangedata}.
\vspace{0.5cm}

\section*{Author contributions}
C.H.Y. conceived the initial experiment concept. H.G.S and H.H.K. designed, performed, and analyzed the measurements. G.D.C and F.B. designed and installed the experimental setup. P.J.S, M.H, J.A.G. and W.D.O. supervised the project.


\appendix
\section{\label{sec:a_derivation} Derivation of the fingerprint pattern}

In this section, we show that the pulse sequence produces a sinusoidal oscillation in the singlet probability as the exchange amplitude is varied, under a quasistatic Gaussian distribution of the nuclear magnetic field gradient $\Delta B_z$. This sinusoidal response gives rise to the fingerprint pattern observed when varying the plunger and barrier pulse amplitudes.

\begin{figure}[ht]
    \centering
    \includegraphics[width=0.3\textwidth]{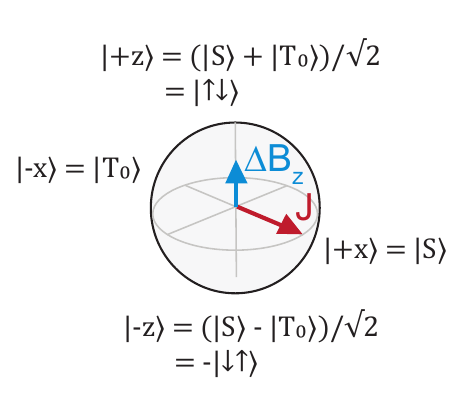}
    \caption{
    \textbf{Two-spin Bloch sphere.} $\ket{S}$ - $\ket{T_0}$ Bloch sphere used in this work. This sphere has been rotated relative to the standard singlet-triplet representation, with $\Delta B_z$ aligned along the z-axis, so that phase accumulation, and the resulting dephasing, during free-evolution is represented naturally as precession about z.
}
    \label{fig:figure1s}
\end{figure}

First, we define the Bloch sphere, as shown in Fig.~\ref{fig:figure1s}, such that
\begin{eqnarray}
\ket{+x} &=& \ket{S},\\
\ket{-x} &=& \ket{T_0},\\
\ket{+z} &=& (\ket{S}+\ket{T_0})/\sqrt{2} = \ket{\uparrow\downarrow} ,\\
\ket{-z} &=& (\ket{S}-\ket{T_0})/\sqrt{2} = -\ket{\downarrow\uparrow},
\end{eqnarray}

In this basis, the Hamiltonian for the double dot system is
\begin{equation}
H
=
\frac{\hbar}{2}
\left(
\omega_{B_z}\sigma_z
-
\omega_J\sigma_x
\right),\label{eq:ham}
\end{equation}
where
\begin{eqnarray}
\omega_{B_z} &=& \frac{g\mu_{\text{B}}\Delta B_z}{\hbar},\\
\omega_J &=& \frac{J}{\hbar}.
\end{eqnarray}

Note that we are assuming quasistatic approximation where $\Delta B_z$ is set to be a constant throughout the entire sequence.

Assuming that the exchange interaction is negligible during the two free-evolution periods (i.e., negligible residual exchange),
\begin{eqnarray}
U_0(t)
&=&
\exp\left(
-i\frac{\omega_{B_z}t}{2}\sigma_z
\right) \nonumber \\
&=&
\begin{pmatrix}
e^{-i\omega_{B_z}t/2} & 0\\
0 & e^{+i\omega_{B_z}t/2}
\end{pmatrix}.
\end{eqnarray}

During the exchange pulse,
\begin{equation}
U_p(t)
=
\exp\left[
-i\frac{t}{2}
\left(
\omega_{B_z}\sigma_z
-
\omega_{J,\text{on}}\sigma_x
\right)
\right],
\end{equation}
where $\omega_{J,\text{on}}=J_\text{on}/\hbar$ and $J_\text{on}$ is the exchange energy during the pulse.

Defining
\begin{equation}
\Omega
=
\sqrt{\omega_{B_z}^2+\omega_{J,\text{on}}^2},
\end{equation}
the unitary for the pulse duration $\tau_{\text{p}}$ becomes

\begin{widetext}
\begin{equation}
U_p(\tau_{\text{p}})
=
\begin{pmatrix}
\cos\left(\frac{\Omega\tau_{\text{p}}}{2}\right)
-i\frac{\omega_{B_z}}{\Omega}
\sin\left(\frac{\Omega\tau_{\text{p}}}{2}\right)
&
+i\frac{\omega_{J,\text{on}}}{\Omega}
\sin\left(\frac{\Omega\tau_{\text{p}}}{2}\right)
\\
+i\frac{\omega_{J,\text{on}}}{\Omega}
\sin\left(\frac{\Omega\tau_{\text{p}}}{2}\right)
&
\cos\left(\frac{\Omega\tau_{\text{p}}}{2}\right)
+i\frac{\omega_{B_z}}{\Omega}
\sin\left(\frac{\Omega\tau_{\text{p}}}{2}\right)
\end{pmatrix}.
\end{equation}

The total evolution for the sequence with idle time $\tau$ and exchange pulse duration $\tau_{\text{p}}$ is therefore
\begin{equation}
U_{\rm tot}
=
U_0(\tau)U_p(\tau_{\text{p}})U_0(\tau)
=
\begin{pmatrix}
e^{-i\omega_{B_z}\tau}
\left[
\cos\left(\frac{\Omega\tau_{\text{p}}}{2}\right)
-i\frac{\omega_{B_z}}{\Omega}
\sin\left(\frac{\Omega\tau_{\text{p}}}{2}\right)
\right]
&
+i\frac{\omega_{J,\text{on}}}{\Omega}
\sin\left(\frac{\Omega\tau_{\text{p}}}{2}\right)
\\
+i\frac{\omega_{J,\text{on}}}{\Omega}
\sin\left(\frac{\Omega\tau_{\text{p}}}{2}\right)
&
e^{+i\omega_{B_z}\tau}
\left[
\cos\left(\frac{\Omega\tau_{\text{p}}}{2}\right)
+i\frac{\omega_{B_z}}{\Omega}
\sin\left(\frac{\Omega\tau_{\text{p}}}{2}\right)
\right]
\end{pmatrix}.
\end{equation}
\end{widetext}

Starting from the singlet state,
\begin{equation}
\ket{\psi_i}=\ket{S}=\frac{1}{\sqrt2}
\begin{pmatrix}
1\\
1
\end{pmatrix},
\end{equation}
the measured singlet probability is
\begin{equation}
P_S
=
\frac{1+\langle\sigma_x\rangle}{2},
\end{equation}
where
\begin{eqnarray}
\langle\sigma_x\rangle
&=& \langle \psi_i|U^\dagger_\text{tot} \hat{\sigma}_x U_\text{tot}|\psi_i\rangle \nonumber\\
&=&
\cos(\Omega\tau_{\text{p}})\cos(2\omega_{B_z}\tau)
\nonumber\\
&&+
\left[
1-\cos(\Omega\tau_{\text{p}})
\right]
\frac{\omega_{J,\text{on}}^2}{\Omega^2}
\cos^2(\omega_{B_z}\tau)
\nonumber\\
&&-
\frac{\omega_{B_z}}{\Omega}
\sin(\Omega\tau_{\text{p}})
\sin(2\omega_{B_z}\tau).
\label{eq:sx_general}
\end{eqnarray}
Equation \ref{eq:sx_general} explicitly shows that the projection onto the
$x$ axis, and hence the singlet probability, undergoes oscillations as
the generalized Rabi frequency during the exchange pulse is varied.

In the ideal limit where the magnetic field gradient is negligible during the
exchange pulse ($\omega_{J,\text{on}} \gg \omega_{B_z}$), we simplify using
\begin{equation}
\omega_{B_z}\rightarrow0
\qquad
\end{equation}
during exchange pulse. Then the exchange unitary reduces to
\begin{equation}
U_p(\tau_{\text{p}})
=
\begin{pmatrix}
\cos\left(\frac{\omega_{J,\text{on}}\tau_{\text{p}}}{2}\right)
&
+i\sin\left(\frac{\omega_{J,\text{on}}\tau_{\text{p}}}{2}\right)
\\
+i\sin\left(\frac{\omega_{J,\text{on}}\tau_{\text{p}}}{2}\right)
&
\cos\left(\frac{\omega_{J,\text{on}}\tau_{\text{p}}}{2}\right)
\end{pmatrix},
\end{equation}
while the free-evolution unitaries remain unchanged. The resulting singlet
projection simplifies to
\begin{eqnarray}
\langle\sigma_x\rangle
&=&
\cos^2(\omega_{B_z}\tau)
-
\sin^2(\omega_{B_z}\tau)
\cos(\omega_{J,\text{on}}\tau_{\text{p}}),
\\
P_S
&=&
\frac12
\left[1+\cos^2(\omega_{B_z}\tau)-\sin^2(\omega_{B_z}\tau)\cos(\omega_{J,\text{on}}\tau_{\text{p}})\right] \nonumber \\
&=&
\frac34+\frac14\cos(2\omega_{B_z}\tau) \nonumber \\ &&-\frac14\left[1-\cos(2\omega_{B_z}\tau)\right]\cos(\omega_{J,\text{on}}\tau_{\text{p}}).
\end{eqnarray}

\textbf{ \begin{figure*}[ht]
            \centering\includegraphics[width=1\textwidth]{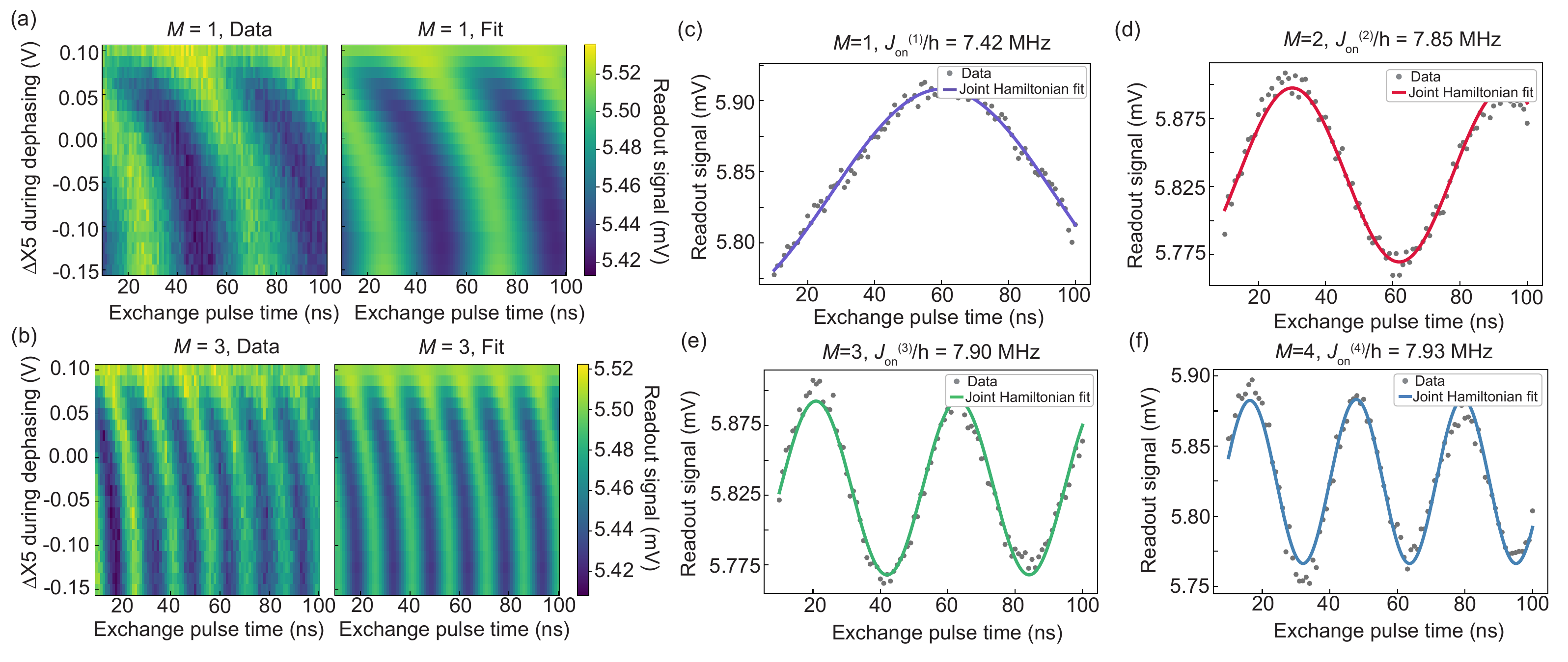} 
            \caption{\textbf{Extracting residual exchange with joint fit of Hamiltonian time evolution simulation.}(a) Left-hand colorplot shows experimental data of exchange oscillations for \textit{M}=1 refocusing pulse, against $\Delta$X5 amplitude during the free-evolution period. A Hamiltonian time evolution is performed for each linecut, with the resulting time evolution jointly fit between the \textit{M}=1 (a) and \textit{M}=3 (b) data. The resulting joint fit, which is used to extract the residual exchange during the free-evolution period is shown in the right-hand plot of (a) and (b). (c) Raw data and evolution simulation for \textit{M}=1 refocusing pulses, with Hamiltonian parameters extracted from a joint fit to the time evolution simulations for \textit{M}=1,2,3,4 refocusing pulses. This data was taken from the star point in the (1,1) cell in Fig.~\ref{fig:figure4}(a). These fits are also shown for (d) \textit{M}=2, (e) \textit{M}=3 and (f) \textit{M}=4 refocusing pulses. The exchange values during the exchange pulse, $J_\text{on}^{(M)}$ are extracted separately for each sequence and shown above each plot. }\label{fig:figure2s}
        \end{figure*}}

Finally, we average the result over many instances of $\Delta B_z$, and hence $\omega_{B_z}$, sampled over multiple experimental repetitions. We model this shot-to-shot variation by a zero-mean Gaussian distribution,
\begin{equation}
p(\omega_{B_z})
=
\frac{1}{\sqrt{2\pi}\sigma_{B_z}}
\exp\left[
-\frac{\omega_{B_z}^2}{2\sigma_{B_z}^2}
\right], \label{eq:gaussian}
\end{equation}
where $\sigma_{B_z}$ is the standard deviation of $\omega_{B_z}$. Once this quasistatic distribution is specified, the ensemble average does not require assuming any particular temporal correlation function or power spectral density for $\Delta B_z$.
Note that temporal correlations of $\Delta B_z(t)$ determine how $\sigma_{B_z}^2$ is related to the underlying noise process, especially for finite-bandwidth or $1/f$-like noise. However, under the quasistatic approximation, these correlations do not enter the ensemble average once $\sigma_{B_z}$, and hence $p(\omega_{B_z})$, has been specified.

The averaged singlet probability is
\begin{equation}
\langle P_S \rangle
=
\int d\omega_{B_z}\,p(\omega_{B_z})P_S(\omega_{B_z}).
\end{equation}
Using
\begin{equation}
\langle \cos(2\omega_{B_z}\tau)\rangle
=
e^{-2\sigma_{B_z}^2\tau^2},
\end{equation}
we obtain
\begin{eqnarray}
\langle P_S \rangle
&=&
\frac{3}{4}
+
\frac{1}{4}e^{-2\sigma_{B_z}^2\tau^2}
\nonumber\\
&&-
\frac{1}{4}
\left[
1-e^{-2\sigma_{B_z}^2\tau^2}
\right]
\cos(\omega_{J,\text{on}}\tau_{\text{p}}).
\end{eqnarray}

For free-evolution times much longer than $T_2^*$, equivalently
\begin{equation}
\sigma_{B_z}\tau\gg1,
\end{equation}
the exponential factor is suppressed. In this fully dephased limit, the averaged singlet probability reduces to
\begin{equation}
\left\langle P_S \right\rangle
=
\frac{3}{4}
-
\frac{1}{4}\cos(\omega_{J,\text{on}}\tau_{\text{p}}).
\end{equation}

Therefore, averaging over quasistatic fluctuations in $\Delta B_z$ yields a characteristic fingerprint pattern as a sinusoidal oscillation of the singlet probability with the exchange pulse area $\omega_{J,\text{on}}\tau_{\text{p}}$. In the fully ensemble-dephased limit, the singlet probability oscillates between $1/2$ and $1$. This ideal contrast, however, relies on the quasistatic approximation. Under realistic conditions, at evolution times exceeding the Hahn-echo coherence time, intra-shot fluctuations in $\Delta B_z$ limit the effectiveness of refocusing, as discussed in the main text.

\section{\label{sec:simulation} Simulations to extract residual exchange}
In order to extract the residual exchange from the exchange oscillations shown in Figs.~\ref{fig:figure3} and \ref{fig:figure4}, we simulate the piecewise time evolution and jointly fit measurements obtained with different numbers of exchange pulses: $\textit{M}=1,3$ for Fig.~\ref{fig:figure3} and $\textit{M}=1,2,3,4$ for Fig.~\ref{fig:figure4}. Based on the estimated separation of the polarized triplet states discussed in Sec.~\ref{sec:1}, we neglect the $\ket{T_\pm}$ states and restrict the simulation to the effective $\{\ket{S},\ket{T_0}\}$ (or equivalently, $\{\ket{\uparrow\downarrow},\ket{\downarrow\uparrow}\}$) subspace.

\begin{figure*}[ht]
    \centering
    \includegraphics[width=1\textwidth]{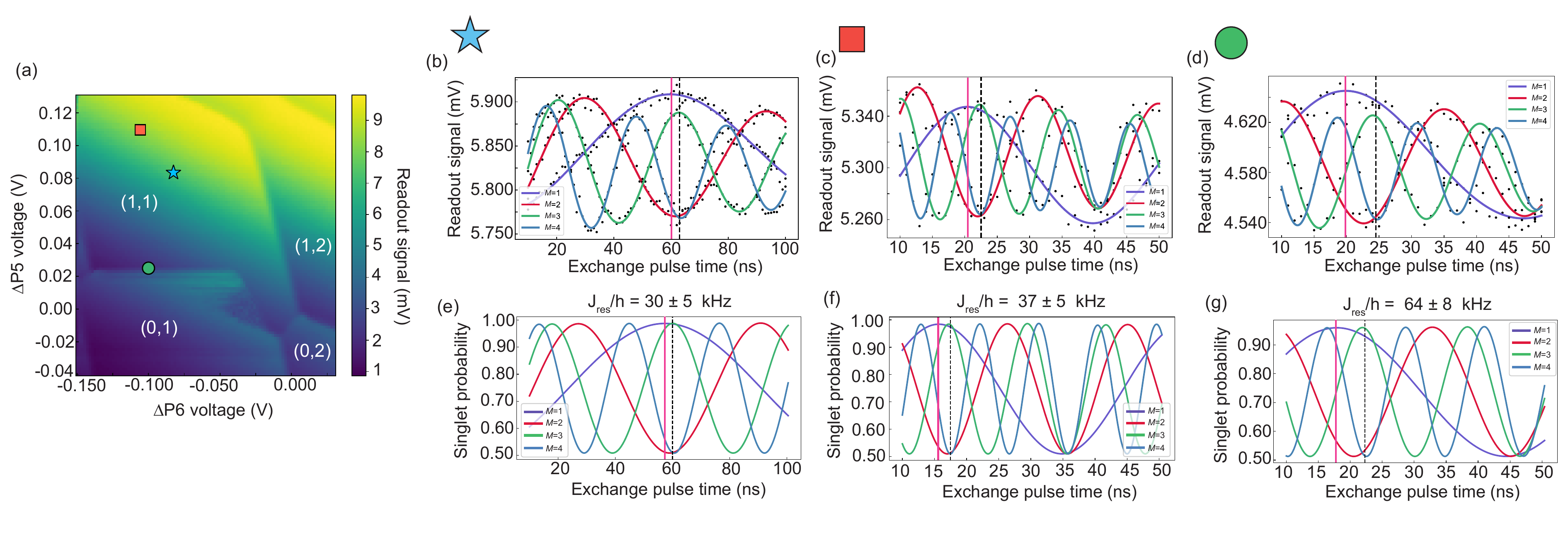}
    \caption{
    \textbf{Time-evolution simulations using the extracted residual exchange.}(a) Charge-stability diagram around the $(1,1)$-$(0,2)$ interdot transition. The blue star, red square, and green circle indicate the free-evolution points used in the subsequent measurements. (b)-(d) Exchange oscillations measured at the blue-star, red-square, and green-circle operating points, respectively, using $M=1$ (purple), $M=2$ (red), $M=3$ (green), and $M=4$ (blue) exchange pulses. The vertical pink line represents the maxima of the $M$=1 curve, while the vertical dashed black line represents the  maxima(minima) of the $M=$3(4) curves. Misalignment between these two vertical lines is a potential indicator of residual exchange at this idle position in the (1,1)  cell. (e)-(g) Corresponding simulated singlet-return probabilities calculated using the fitted values of $J_{\mathrm{res}}$, $J_{\mathrm{on}}^{(M)}$ for the measurements in (b)-(d), respectively. A $\sigma_{B_z}$ value of 140 kHz, extracted from the $M=$1,3 simulations in Fig.\ref{fig:figure2s} was used for these simulations.
}
    \label{fig:figure3s}
\end{figure*}

The simulation starts in the singlet state and applies the sequence containing $M$ pulses:
\begin{equation}
    \text{wait}(\tau)
    -
    \text{pulse}(\tau_{\text{p}})
    -
    \left[
    \text{gap}(\tau_g)
    -
    \text{pulse}(\tau_{\text{p}})
    \right]^{M-1}
    -
    \text{wait}(\tau),
    \label{eq:residual_exchange_sequence}
\end{equation}
where $\tau_g=20~\mathrm{ns}$. In contrast to the idealized treatment in Appendix~\ref{sec:a_derivation}, the exchange interaction during the wait and gap segments is not assumed to vanish. For the wait and gap segments, we define
\begin{equation}
    U_{0'}(t)
    =
    \exp\left[
    -i\frac{t}{2}
    \left(
    \omega_{B_z}\sigma_z
    -
    \omega_{J,\text{res}}\sigma_x
    \right)
    \right].
    \label{eq:U_0prime}
\end{equation}
Here, $\omega_{J,\text{res}}= J_{\text{res}}/\hbar$ is the residual exchange in the unit of angular frequency. We also modify the evolution operator during an exchange pulse as
\begin{equation}
    U_p^{(M)}(t)
    =
    \exp\left[
    -i\frac{t}{2}
    \left(
    \omega_{B_z}\sigma_z
    -
    \omega_{J,\text{on}}^{(M)}\sigma_x
    \right)
    \right],
    \label{eq:U_pM}
\end{equation}
where $\omega_{J,\text{on}}^{(M)}=J_{\text{on}}^{(M)}/\hbar$ is the exchange for the sequence containing $M$ exchange pulses. Allowing $J_{\text{on}}^{(M)}$ to depend on $M$ accounts for pulse distortion in the experimental implementation.

The total evolution operator for a sequence containing $M$ exchange pulses is
\begin{equation}
    U_{\rm tot}^{(M)}
    =
    U_{0'}(\tau)
    \left[
    U_p^{(M)}(\tau_{\text{p}})
    U_{0'}(\tau_g)
    \right]^{M-1}
    U_p^{(M)}(\tau_{\text{p}})
    U_{0'}(\tau).
    \label{eq:U_tot_M}
\end{equation}
For $\textit{M}=1$, this reduces to
\begin{equation}
    U_{\rm tot}^{(1)}
    =
    U_{0'}(\tau)
    U_p^{(1)}(\tau_{\text{p}})
    U_{0'}(\tau),
\end{equation}
which is analogous to the $U_{\rm tot}=U_0U_pU_0$ evolution for the zero-residual-exchange case.

Because the initialized singlet state corresponds to $\ket{S}=\ket{+x}$ in our Bloch-sphere convention, the simulated observable is
\begin{equation}
    \langle\sigma_x\rangle^{(M)}
    =
    \bra{S}
    \left(U_{\rm tot}^{(M)}\right)^\dagger
    \sigma_x
    U_{\rm tot}^{(M)}
    \ket{S}.
    \label{eq:sigma_x_M}
\end{equation}
The corresponding singlet-return probability is
\begin{equation}
    P_S^{(M)}
    =
    \frac{1+\langle\sigma_x\rangle^{(M)}}{2}.
    \label{eq:PS_M}
\end{equation}

Following Eq.~\ref{eq:gaussian}, we assume $\omega_{B_z}$ to be a Gaussian random variable. From the joint fit over the Hamiltonian time evolution simulation, we extract a $\sigma_{B_z}$ value of 140 kHz. Then Gauss-Hermite quadrature is used to average the simulated singlet-return probability over this distribution,
\begin{equation}
    \left\langle P_S^{(M)} \right\rangle
    =
    \int d\omega_{B_z}\,
    p(\omega_{B_z})
    P_S^{(M)}(\omega_{B_z}).
    \label{eq:PS_average}
\end{equation}

In addition to non-zero residual exchange $J_\text{res}$ and $M$-dependent exchange during pulse $J_\text{on}^{(M)}$, we also allow a fixed pulse-time offset, which adds a fixed phase shift, to be fit jointly for the four datasets, to account for finite experimental pulse rise time.


Because the experimental data was not obtained in single-shot mode, every linecut is mapped onto the simulation using an independently fitted affine transformation: signal = offset + contrast $\times$ singlet probability. This allows the Hamiltonian parameters to be extracted from the phase and frequency of the exchange oscillations, without requiring identical readout amplitude and offset for all measurements. Uncertainties were estimated from the optimizer's Jacobian.

Figure~\ref{fig:figure2s}(a,b) shows the results of this simulation for the \textit{M}=1 and \textit{M}=3 refocusing pulse data shown in Fig.~\ref{fig:figure3}(b,d). Each linecut is simulated and jointly fit between the \textit{M}=1 and \textit{M}=3 data, as described above, to extract a residual exchange value for each X5 amplitude, as shown  Fig.~\ref{fig:figure3}(g). The resulting fits for each X5 amplitude are plotted as a 2D colorplot for both sequences, showing good correspondence with measured data.

Figure~\ref{fig:figure2s}(c-f) shows the raw data and example simulation fit results for the \textit{M}=1,2,3,4 pulse sequences measured at the star point in the (1,1) cell, shown in Fig.~\ref{fig:figure4}(f). Extracted $J_\text{on}^{(M)}$ values are displayed on top of each plot, with a $J_\text{res}/h$ value of 30 $\pm$ 5 kHz being extracted from the joint fit to all datasets. 

As a consistency check, we use this simulation code to extract the $J_\text{res}$ and $J_\text{on}^{(M)}$ values for the exchange oscillation data taken at the star [Fig.~\ref{fig:figure3s}(b)], square [Fig.~\ref{fig:figure3s}(c)] and circle [Fig.~\ref{fig:figure3s}(d)] point in the (1,1) cell, shown in Fig.~\ref{fig:figure3s}(a). We then enter these extracted values, along with our extracted value of $\sigma_{\Delta_{B_z}}$ = 140 kHz, into a different simulation code, which also carries out a Hamiltonian time evolution of each sequence, but this time using Monte Carlo to sample $\omega_{B_z}$ values over the magnetic gradient distribution. This code does not fit or extract any values but allows us to visually compare the time evolution to experimental data. These simulated time evolutions using our extracted Hamiltonian parameters, for each free-evolution point in the (1,1) cell, are shown in Figs.~\ref{fig:figure3s}(e-g) and show good agreement with experimental data [Figs.~\ref{fig:figure3s}(b-d)]. We attribute the phase offset observed between the simulated time evolution and experimental data to the finite rise time of the pulses, which is not included in this time evolution simulation. The effect of this finite rise time is more pronounced for higher $J_\text{on}^{(M)}$ values as expected.

\section{\label{sec:init} Initialization optimization}

 \begin{figure*}[ht]
            \centering\includegraphics[width=1\textwidth]{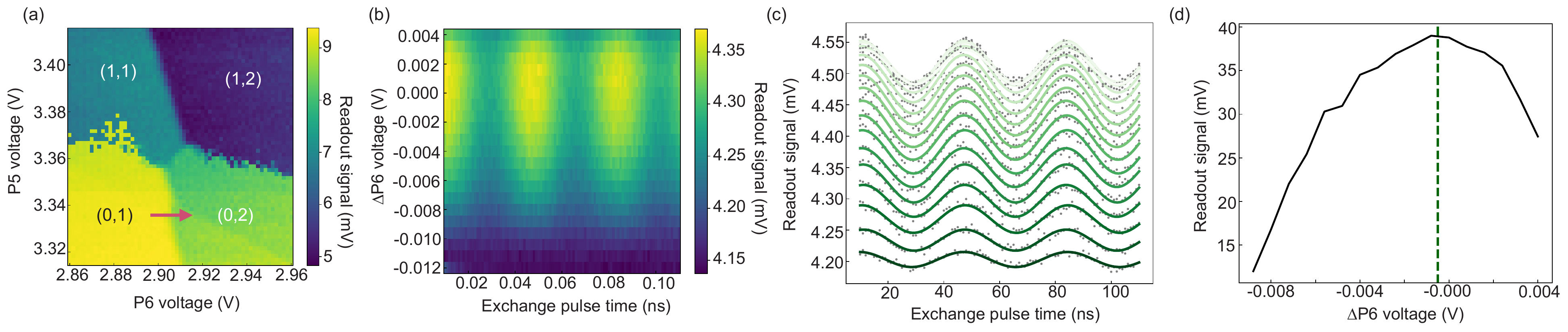} 
            \caption{\textbf{Exchange oscillations for singlet initialization optimization.} (a) Charge stability map around the (1,1)-(0,2) interdot transition. The pink arrow indicates the P6 initialization sweep across the (0,1)-(0,2) transition.(b) Exchange oscillations as a function of P6 initialization voltage, for a fixed P5 voltage. P6 is swept across the (0,1)-(0,2) transition as indicated by the pink arrow in (a). (c) Vertically offset waterfall plot of the data shown in (b), for clarity. (d) Exchange oscillation amplitude as a function of P6 initialization voltage, extracted from fits to the oscillations in (b),(c). Vertical dashed line denotes the optimal P6 initialization voltage that maximises readout visibility. }\label{fig:figure4s}
        \end{figure*}

We use the visibility of exchange oscillations to identify an optimal singlet-initialization point in gate-voltage space.

Figure~\ref{fig:figure4s}(a) shows a charge-stability map centered on the (1,1)-(0,2) transition used for the Pauli spin blockade readout. To initialize the singlet, we set the P5 and P6 voltages close to the (0,1)-(0,2) charge transition and wait for an initialization time $\tau_{\text{init}}$
 that exceeds the relaxation time $T_1$ at this location \cite{bluhm2010enhancing}. Figures~\ref{fig:figure4s}(b,c) show exchange-oscillation measurements acquired while varying the P6 initialization voltage at a fixed P5 initialization voltage. The P6 voltage is swept across the (0,1)-(0,2) transition along the trajectory indicated by the pink arrow in Fig.~\ref{fig:figure4s}(a).

Figure~\ref{fig:figure4s}(d) shows the extracted exchange oscillation amplitudes from Figs.~\ref{fig:figure4s}(b,c) as a function of P6 initialization voltage. From the maximum of this plot, we find an optimal singlet initialization voltage that maximizes readout visibility.

\section{\label{sec:eo} Exchange oscillation properties}

 \begin{figure}[h]
            \centering\includegraphics[width=0.44\textwidth]{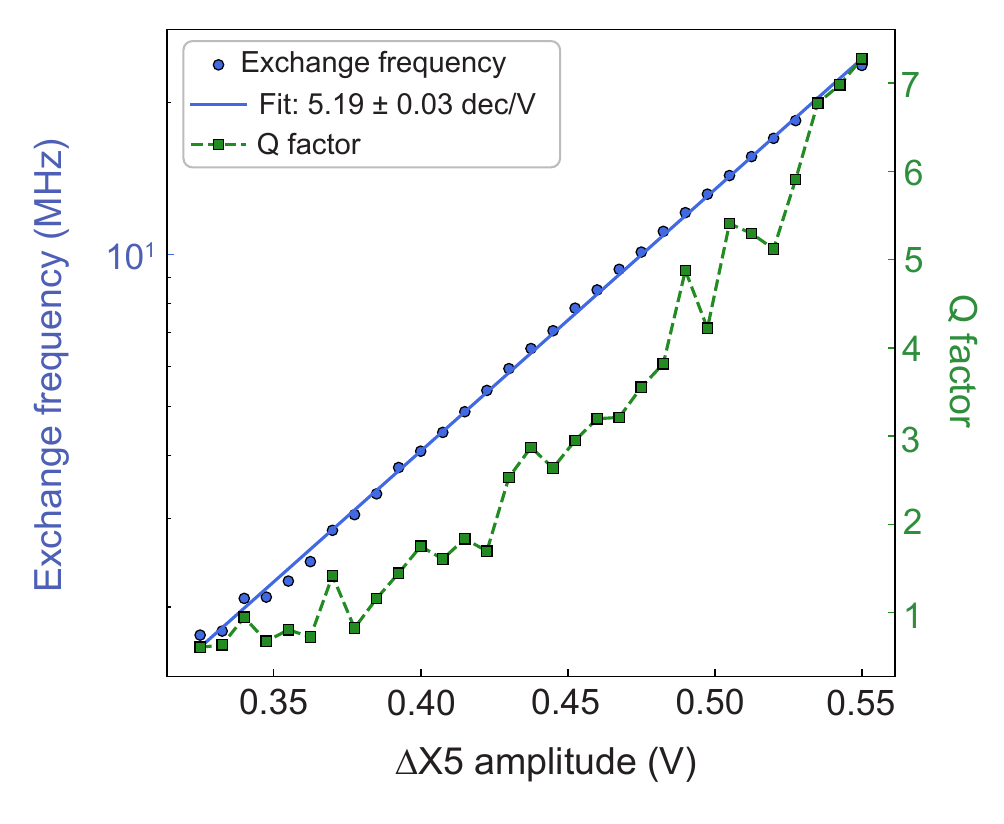} 
            \caption{\textbf{Exchange oscillations parameter extraction.} Dual y-axis plot showing the exchange oscillation frequency extracted from the plot in Fig.~\ref{fig:figure1}(g) (left, blue axis) and the quality factor (Q-factor) of these oscillations (right, green axis). The exchange oscillation frequencies are plotted on a log scale and fit to a linear fit, to extract an exchange tunability from the X5 gate of 5.2 dec/V. The Q-factor is defined as the number of exchange oscillations within the characteristic decay time.}\label{fig:figure5s}
\end{figure}
Figure~\ref{fig:figure5s} shows the exchange oscillation frequency and oscillation quality-factor (Q-factor) extracted from the exchange oscillations as a function of $\Delta$X5 voltage, shown in Fig.~\ref{fig:figure1}(g). From this data we extract an exchange tunability of 5.2~dec/V. 

\newpage

%

\end{document}